\documentclass{article}
\usepackage{spconf,amsmath,graphicx}

\usepackage{makecell}
\usepackage{multirow}
\usepackage{amssymb}
\usepackage{color}
\usepackage{url}
\usepackage[T1]{fontenc}
\usepackage{enumitem}
\usepackage{fancyhdr}
\setlist{nolistsep,leftmargin=1em}
\usepackage{booktabs}   % For prettier tables
\usepackage{array}      % For centered cells with a specified width

\usepackage{tabularx}
\usepackage[dvipsnames]{xcolor}

\newcommand{\para}[1]{\noindent{\bf{#1}}}
\usepackage{float}
\usepackage{listings}

\newcommand{\av}[1]{\textcolor{black}{#1}}
\newcommand{\ao}[1]{\textcolor{black}{#1}}
\newcommand{\samuel}[1]{\textcolor{black}{#1}}
\title{A sound approach: Using large language models to generate audio descriptions for egocentric text-audio retrieval}
\fiveauthors{Andreea-Maria Oncescu$^1$}{Jo{\~a}o F. Henriques$^1$}{Andrew Zisserman$^1$}{Samuel Albanie$^2$}{A. Sophia Koepke$^3$}{$^1$University of Oxford}{$^2$University of Cambridge}{$^3$University of  T{\"u}bingen}

\begin{document}
\maketitle
\begin{abstract}
% In this study we propose two new audio-retrieval benchmarks based on the Epic-Kitchens' Multi-Instance Retrieval (EpicMIR) task and on the EgoMCQ task which is based on Ego4D. Additionally, we introduce a methodology of generating improved audio-centric captions for these datasets by employing Large Language Models (LLMs) such as ChatGPT and GPT-4. We show that this approach provides state-of-the art results on the original benchmarks proposed. We also demonstrate that LLMs can be used to determine the difficulty of identifying the action associated with a sound when no other modality (e.g.\ video) is available. Lastly, we show that jointly evaluating zero-shot audio and video retrieval models on EpicMIR improves video retrieval results by more than 2\%.

%%%New abstract draft that tries to address Joao's comment:
\noindent Video databases from the internet are a valuable source of text-audio retrieval datasets. 
However, given that sound and vision streams represent different ``views'' of the data, treating visual descriptions as audio descriptions is far from optimal. \ao{Even if audio class labels are present, they commonly are not very detailed, making them unsuited for text-audio retrieval.}
To exploit relevant audio information from video-text datasets, we introduce a methodology for generating audio-centric descriptions using Large Language Models (LLMs).
In this work, we consider the egocentric video setting and propose \ao{three} new text-audio retrieval benchmarks based on the EpicMIR and EgoMCQ tasks, \ao{and on the EpicSounds dataset}. 
Our approach for obtaining audio-centric descriptions gives significantly higher zero-shot performance than using the original visual-centric descriptions. 
\ao{Furthermore, we show that using the same prompts, we can successfully employ LLMs to improve the retrieval on EpicSounds, compared to using the original audio class labels of the dataset. Finally, we confirm that LLMs can be used to determine the difficulty of identifying the action associated with a sound.}
\end{abstract}

\begin{figure*}
    \centering
\includegraphics[trim=0 25 0 0, clip,width=1.0\textwidth]{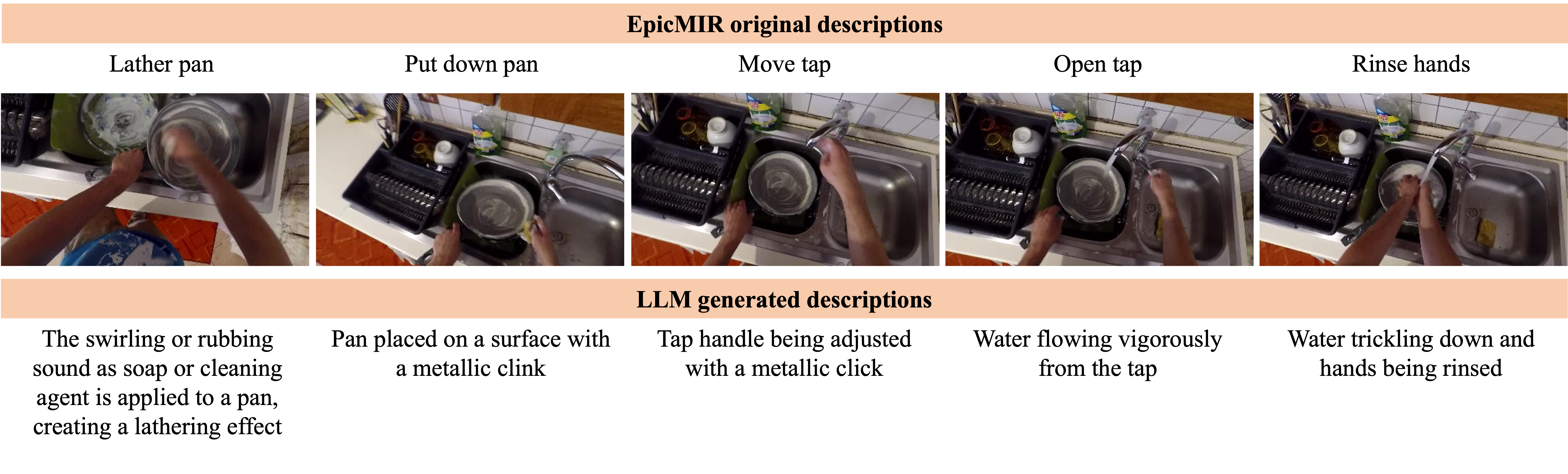}
\vspace{-1.5em}
    \caption{Frames from the EpicKitchens dataset together with the corresponding original visual descriptions from EpicMIR shown above and those generated with our approach (using the ChatGPT LLM~\cite{chatgpt}) below.}
    \label{fig:img1}
    \vspace{-1em}
\end{figure*}

\let\thefootnote\relax\footnotetext{Acknowledgements. This work is supported by the EPSRC (VisualAI EP/T028572/1 and DTA Studentship), the Royal Academy of Engineering (RF\textbackslash201819\textbackslash18\textbackslash163), \samuel{an Isaac Newton Trust Grant} and the DFG - EXC 2064/1 - project  390727645. We are grateful
to Jaesung Huh, Triantafyllos Afouras and Bernie Huang for their helpful comments and suggestions.}

\section{Introduction}
\label{sec:intro}
Searching the ever-expanding supply of audio and video media hosted online has become a key technical challenge.
Concurrently, LLMs have become more powerful, exhibiting early signs of commonsense reasoning and primitive world modelling~\cite{2023arXiv230709288T}.
Given their extensive text-based knowledge about the sensory world, in this work we ask whether LLMs can improve search capabilities for other modalities such as audio and video.
% However, the introduction of LLMs has resulted in significant improvements on this task in the text domain.
% we explore the potential for LLMs to aid more effective searching through modalities other than natural language.
In particular, we consider the task of egocentric audio retrieval from text queries. \par

% In the rapidly evolving landscape of multimedia data, the fusion of diverse modalities has opened up exciting avenues for enhancing user experiences and enabling new applications. Audio and visual content, two pivotal sources of information, offer distinct perspectives on the world around us. However, harnessing the complementary nature of these modalities for content retrieval remains a challenge. While tremendous strides have been made in both visual and audio recognition, the seamless integration of these modalities is yet to be fully realized.

% One avenue for exploiting LLMs for audio data is to tap into the vast general knowledge these models hold.
% In the multimodal setting, this can be done for example by training small modules on top of pre-trained frozen models~\cite{li2023blip2} to align the text embeddings (from the LLMs) and image/audio/video embeddings.
Our strategy is to employ text as an intermediate medium for aligning vision and audio signals by leveraging the text-based knowledge that an LLM possesses about sight and sound.
Concretely, we use LLMs to generate plausible {\em audio} descriptions for videos \av{when given their {\em visual} descriptions.}
To do so, we leverage the LLM's in-context learning capability,
%ChatGPT~\cite{chatgpt}, 
and provide it with exemplars of the desired mapping -- pairs of visual-centric and corresponding audio-centric descriptions.
This few-shot shot approach is made possible by the existence of a small collection of content that has been annotated with both visual-centric and audio-centric descriptions.
In this work, we source these pairs from overlapping samples between the Kinetics700-2020~\cite{Smaira2020ASN} and AudioCaps~\cite{kim2019audiocaps} datasets (we refer to these examples as $\text{Kinetics} \cap \text{AudioCaps}$). 
Fig.~\ref{fig:img1} shows examples of few-shot generated audio descriptions produced with our approach. \par
% \ask{Elaborate on the kinds of descriptions that are generated, maybe briefly mention how you obtain those, since this is the main contribution.}
With this ``converter'' in hand, we scale its application to the conversion of full text-video datasets to text-audio datasets. % from existing video datasets.
% \ask{But this is not what you do, is it? You only report zero-shot results... In that case, maybe we should not mention this here.} \par
Specifically, we construct two text-audio datasets derived from egocentric video retrieval tasks sourced from EpicKitchens~\cite{Damen2018EPICKITCHENS} and Ego4D~\cite{ego4d}, and demonstrate the value of this data empirically. \ao{We additionally apply} \samuel{a similar methodology} \ao{to the {\em audio class labels} of  EpicSounds to improve retrieval results.}
We also demonstrate that LLMs can usefully predict when the action within a video can be reliably determined solely from its audio track, with applications for curating new text-audio datasets.
\vspace{-5pt}
% This insight potentially enables guidance for curating new text-audio datasets, particularly by eliminating text-audio pairs that might introduce noise.\par
% it could aid in decision-making about when to incorporate the audio modality for multimodal video retrieval. \ask{If you hint at this, you should also comment on your results for this,...} 

% Additionally, we explore the possibility of using LLMs to identify when the action/object source of the audio track corresponding to a video can be confidently inferred just by listening to the audio. This can be useful for deciding when to use the audio modality in the context of multimodal video retrieval. It can also be used when collecting new audio-text datasets by removing noisy text-audio pairs.\par

% The ubiquity of Large Language Models, empowered by recent advancements in artificial intelligence, presents a transformative opportunity to bridge the gap between visual and audio descriptions. These models, pre-trained on vast and diverse datasets, possess the capability to comprehend and generate human-like textual descriptions of content across various modalities. By leveraging the semantic richness embedded within these models, we can establish a more comprehensive and intuitive connection between visual and audio elements, revolutionizing the way we search and retrieve multimedia content.

% \ask{I think it would be good to mention those before already, maybe you can add one paragraph about the proposal of the benchmarks}
\vspace{-3mm}
\section{Related work}
\label{sec:relatedwork}

\para{Text-audio retrieval and LLMs.}
Text-audio retrieval entails searching for the most appropriate audio file for a given textual query.
This task was popularised by~\cite{Oncescu21a,koepke2022audio} (though related themes were studied previously~\cite{slaney2002semantic}), and has seen recent improvements through the use of transformer-based models~\cite{laionclap2023}. 
In adjacent fields, there has been a surge of efforts that harness LLMs, e.g.\ GPT-4~\cite{OpenAI2023GPT4TR}, Vicuna~\cite{vicuna2023}, and Llama~\cite{touvron2023llama,2023arXiv230709288T}, for multimodal tasks that require vision-language~\cite{zhu2023minigpt,shen2023hugginggpt,Kaul23,menon2022visual} and audio-language understanding~\cite{mei2023wavcaps}.
\cite{menon2022visual} and \cite{Kaul23} demonstrate the benefits of using LLM-generated textual class descriptions for zero-shot image classification and open-vocabulary object detection respectively.
More closely related to our approach,~\cite{mei2023wavcaps} employ ChatGPT~\cite{chatgpt} for standardizing audio descriptions across various audio-centric datasets.
These refined text-audio pairs are then used to pretrain text-audio models. 
In contrast to prior work which focuses predominantly on audio-centric datasets (as characterized by the availability of audio descriptions or labels), we focus on leveraging datasets that \ao{generally} possess only visual-centric descriptions.

\para{Egocentric audio-visual understanding.}
While most video datasets are captured from a third-person perspective, recent research has shifted towards egocentric data, filmed from a first-person viewpoint. 
The egocentric datasets, EpicKitchens~\cite{Damen2018EPICKITCHENS} and Ego4D~\cite{ego4d}, have been used for various tasks, such as action recognition~\cite{NEURIPS2022_96471570, Planamente_2022_WACV}, moment localization and video retrieval~\cite{lin2022egocentric,zhao2023lavila,Ashutosh_2023_CVPR,pramanick2023egovlpv2} with a focus on the video stream.
% In contrast, we specifically consider their audio tracks.
We, however, focus specifically on their audio tracks.
\par
%\par
In a similar vein to~\cite{Kazakos2019EPICFusionAT}, we analyse how to best exploit the audio information in the egocentric video setting.
% Also related, recent work introduced new audio classification benchmarks on the EpicKitchens dataset~\cite{EPICSOUNDS2023}.
Also related, recent work introduced EpicSounds~\cite{EPICSOUNDS2023}, an audio classification benchmark on the EpicKitchens~\cite{Damen2018EPICKITCHENS} dataset.
% \cite{Kazakos2019EPICFusionAT} showed that the recorded audio can be noisy at times, due to the recording procedure. However, even when noisy, it is shown that the audio modality still contains useful cues. 
In contrast to these approaches, we consider the retrieval task rather than classification and
%
%In this work the authors have manually annotated all the audio content existent in the Epic-Kitchens dataset. 
%They have benchmarked the newly collected data on the task of audio classification. 
employ LLMs to automatically generate additional audio descriptions for text-audio retrieval.\par
\vspace{-8pt}
\section{Datasets and approach}
\label{sec:method_data}
We first summarise existing relevant datasets in Sec.~\ref{sec:existing_datasets}. Next, we detail our proposed method for employing Large Language Models (LLMs) in two key areas: firstly, to bridge the gap between visual and audio descriptions; \ao{and secondly, to create audio descriptions from audio labels}. This approach is aimed at improving text-audio retrieval in egocentric datasets, as discussed in Sec.~\ref{sec:generating_audio_desc}.
We introduce our AudioEpicMIR, AudioEgoMCQ \ao{and EpicSoundsRet} benchmarks in Sec.~\ref{sec:tasks_datasets}.  
Lastly, we present our method for evaluating the level of informativeness of audio samples in Sec.~\ref{subsec:llmssplits}.
\vspace{-5mm}
\subsection{Datasets and tasks}\label{sec:existing_datasets}
\noindent\textbf{EpicKitchens~\cite{Damen2018EPICKITCHENS} \& EpicMIR~\cite{damen2022rescaling}}. EpicKitchens contains 100 hours of recordings filmed from first-person perspective. The Multi-Instance Action Retrieval (EpicMIR)~\cite{damen2022rescaling} task based on EpicKitchens consists of finding the most relevant video given a text query, and vice-versa. The test set contains 9,668 videos and corresponding visual descriptions in the form \texttt{verb/s $+$ noun/s}. 
 %We extract the audio files in mono format together with their original visual descriptions.
 The average number of words per sentence is 2.93 with standard deviation 1.17.

\noindent\textbf{Ego4D~\cite{ego4d} \& EgoMCQ~\cite{lin2022egocentric}}. Ego4D is the largest egocentric dataset with over 3,670 video hours and accompanying narrations.
%This is also more diverse than EpicKitchens in terms of filming locations. 
%Recorded videos are split into clips with given start and end time. Clips also contain narrations that can be used for retrieval. 
The EgoMCQ~\cite{lin2022egocentric} task based on Ego4D consists of 39,751 pairs containing one description and 5 video clips each.
It aims at
finding the correct clip for a given description.
% The task is split into two smaller tasks, the inter-video task where the 5 clips are taken from different videos, and the more challenging intra-video task where the 5 clips are taken from the same video. 
\par
\noindent\ao{\textbf{EpicSounds~\cite{EPICSOUNDS2023}} is sourced from the EpicKitchens dataset. It contains only those audio tracks that are useful for the audio classification task as verified through manual annotations.
%process Its creation included annotators going through all the EpicKitchens video soundtracks, choosing only those audio parts that are useful for the task of audio classification. 
44 different classes are labelled for 10,276 test audio chunks.
}\par
\noindent\textbf{Kinetics700-2020~\cite{Smaira2020ASN}}  contains 10s clips from YouTube and corresponding activity labels which are in the form \texttt{verb/s + noun/s}.
The average word count is
2.09 with a standard deviation of 0.79.\par
\noindent\textbf{AudioCaps~\cite{kim2019audiocaps}} is curated from YouTube and contains 10s audio files and text descriptions. It serves as a common benchmark dataset for audio captioning and text-audio retrieval.

\vspace{-3mm}
\subsection{Audio description generation methodology}\label{sec:generating_audio_desc}
\begin{figure}
    \centering
    \includegraphics[scale=0.35]{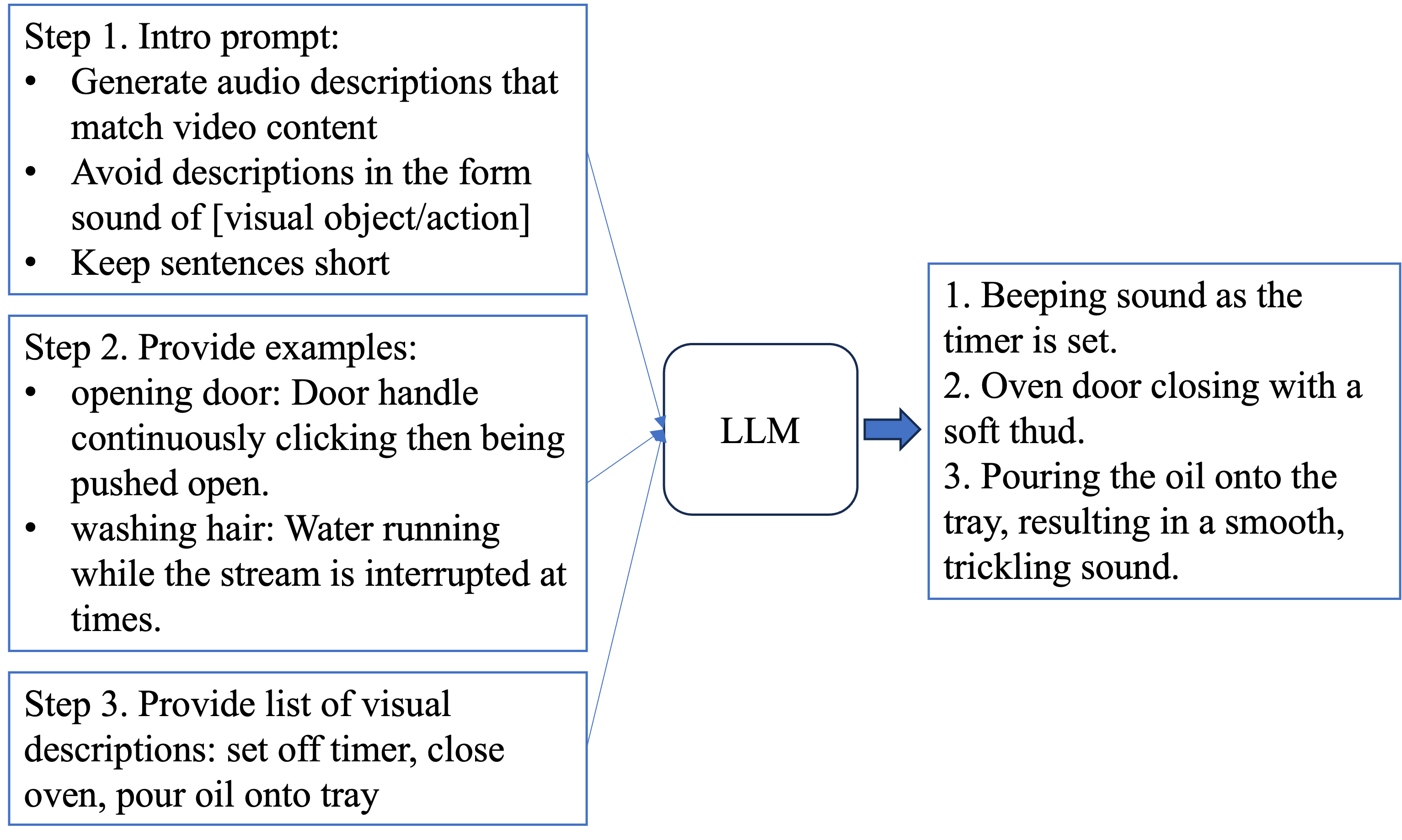}
   % \vspace{-1em}
    \caption{Given visual-centric descriptions, we propose to use an LLM (ChatGPT) to generate audio descriptions (\textit{step 3}).
    The LLM is prompted with a task description (\textit{step 1}) and few-shot paired examples of visual-centric and audio descriptions (\textit{step 2}).}
    \label{fig:chatgpt_prompts}
    % \vspace{-10pt}
\end{figure}

% \ask{Given that this is the main contribution, this should probably be expanded. Maybe you can provide some concrete example prompt if it's not too long. Also, are the final descriptions obtained with a two-stage generation protocol?}
% An initial observation revealed that the visual descriptions for Kinetics700-2020~\cite{Smaira2020ASN} clips bear stylistic similarities to those in EpicKitchens and more relevant to our case, to those in EpicMIR. These is because the Kinetics descriptions are of similar format and length as those in EpicMIR.
% Our approach leverages ChatGPT for the audio tracks of the videos from EpicMIR and EgoMCQ, we leverage ChatGPT. 
As noted in Sec.~\ref{sec:intro}, we find that there are a number of clips in common between Kinetics700-2020~\cite{Smaira2020ASN} and AudioCaps~\cite{kim2019audiocaps} (283 in total).
We use example correspondences to condition the LLM to generate audio descriptions in the style of AudioCaps given access to visual verb-noun descriptions. 
This approach is shown in Fig.~\ref{fig:chatgpt_prompts}, where we first give a general task description, together with heuristic constraints that were obtained by manual experimentation on a handful of samples (a form of ``prompt engineering'').

% These \textit{sound of [object/action]} or \textit{keeping the description shorter} (step 1). 

We combine these instructions with few-shot paired examples of visual descriptions and audio descriptions sampled from $\text{Kinetics} \cap \text{AudioCaps}$.
In particular, we select 14 pairs to balance sufficient examples while preserving the model's focus on the task prompt (step 2). 
Finally, we provide the LLM with the visual descriptions \ao{or audio labels} for which we want to generate new audio descriptions (step 3).
We developed this approach by experimenting with samples from EpicMIR, and apply the same strategy directly to EgoMCQ \ao{and EpicSounds} in Sec.~\ref{sec:experiments}.
% As a result, the sample description pairs are predominantly quite different from the EgoMCQ descriptions in terms of style and length. 
% \begin{figure}[htb]
%     \centering
%     \includegraphics[scale=0.4]{images/youtube_ids_all_kin_ac.png}
%     \caption{Diagram of common Kinetics and AudioCaps clips}
%     \label{fig:kin_ac}
% \end{figure}

\vspace{-3mm}
\subsection{Our proposed text-audio retrieval benchmarks}\label{sec:tasks_datasets}
We apply our approach for generating audio descriptions to \ao{EpicMIR, EgoMCQ, and EpicSounds}. As a result, we obtain \ao{three} new benchmarks, namely AudioEpicMIR, AudioMCQ, \ao{and EpicSoundsRet}. We refer to the original visual descriptions/audio class labels for all benchmark datasets as \textit{Aud orig} and to our LLM-generated descriptions as \textit{Aud LLM}.

 \noindent\textbf{AudioEpicMIR} is curated from the EpicMIR task 
 %which contains 9668 videos and corresponding visual descriptions in the form \texttt{verb/s $+$ noun/s}. We curate AudioEpicMIR
 by extracting the audio tracks of the EpicMIR videos and keeping their original visual descriptions. Additionally, we generate audio-centric descriptions using ChatGPT (GPT-3.5)~\cite{chatgpt} as the LLM in our methodology, as described in Sec.~\ref{sec:generating_audio_desc}.
 \par
 %  To confirm that EpicMIR and Kinetics700-2020 have indeed similar style descriptions we have calculated the average length and standard deviation of the sentences of the EpicMIR evaluation set and Kinetics validation set.
 % whilst Kinetics has 2.09 with a deviation of 0.79.
 % We then employ ChatGPT to generate AudioCaps style descriptions starting from the original video descriptions.\par
% \input{tables/kin_to_ac_eg}

\noindent\textbf{AudioEgoMCQ} is gathered based on the EgoMCQ task. We observe that in this dataset, some of the videos do not have an audio soundtrack at all.
Therefore, to generate a text-audio dataset, we need to exclude some of the original text-video pairs. Specifically, for the \textbf{intra-video} task, we excluded all text-video pairs when the video corresponding to the text query did not contain sound. For \textbf{inter-video}, we additionally replaced clips without audio from the pairs by clips with audio. Finally, we exclude text-video pairs if any of the five videos have a silent soundtrack. This results in 23,121 text-audios pairs. The remaining original visual descriptions contain an average of 8.15 words (with a standard deviation of 3.01). We use our audio description generation approach described in Sec.~\ref{sec:generating_audio_desc} to obtain audio-centric descriptions. As before, we use ChatGPT (GPT-3.5)~\cite{chatgpt} as the LLM.

\noindent\ao{\textbf{EpicSoundsRet} differs from EpicMIR and EgoMCQ in that it originally contains \textit{audio class labels}, not \textit{visual descriptions}. 
%This benchmark is curated from the EpicSounds dataset. 
For the retrieval task we use the audio class labels as text queries together with the corresponding audio files. 
We use our LLM prompts out of the box to generate \textit{audio descriptions} for EpicSounds starting from the \textit{audio labels}.
% We note however that, for all other experiments in this paper, we generated \textit{audio descriptions} using LLMs starting from \textit{visual} class labels. 
% As a result, our prompts are \textit{not} necessarily optimal for \textit{audio} class labels as inputs. 
}

\vspace{-3mm}
\subsection{Determining the audio relevancy using LLMs}
\label{subsec:llmssplits}
We observe that many video clips contain audio that is too noisy or generic to inform the text-audio retrieval process. An example is `taking clothes from basket' or `putting clothes in basket' which both have similar associated sounds that are hard to differentiate. 
This prompts the question, can we \samuel{identify such audio by looking only} at the given visual description?
% Can language models point us to a subset of video-text descriptions where the audio is more likely to be informative enough to allow for performing text to audio retrieval? \ask{Why do we want to know this? Either add this here, or after describing the categories.} \ask{This did not work though?}
To explore this, we tasked GPT-4 with splitting the original visual text descriptions into three categories according to the relevancy of the sound:
\begin{itemize}
    \item High: audio is very informative for the visual task, e.g.\ `turning on tap', `washing dishes'.
    \item Moderate: audio is informative but not enough information is provided in the text description for it to be identifiable, e.g.\ `putting a plate down' has a different sound based on if it is placed on a kitchen table or a sofa.
    \item Low: audio is not likely to be informative, e.g.\ `get carrot'.
\end{itemize}
% \vspace*{-6pt}
\vspace{-3mm}
\section{Experiments}

\label{sec:experiments}

%Our experiments consider the audio retrieval performance of the best LAION-Clap model checkpoint and of the WavCaps model finetuned on the AudioCaps or Clotho datasets applied zero-shot to the egocentric setting, i.e.\ without any finetuning on egocentric data.
\noindent \textbf{Evaluation metrics.} On AudioEpicMIR and \ao{EpicSoundsRet}, we report Mean Average Precision (mAP) and normalised Discounted Cumulative
Gain (nDCG)~\cite{10.1145/582415.582418} following \cite{damen2022rescaling}. 
nDCG measures the relevance of text descriptions in response to a video query by positively acknowledging semantically analogous descriptions in addition to the correct pair. %The relevancy score is defined based on repeated verbs and nouns. 
For AudioEgoMCQ, we report the Retrieval@1 score, similar to~\cite{lin2022egocentric}. We consider the following two tasks. For the more challenging \textbf{intra-video} task we are given a text query and aim to find the corresponding clip out of 5 clips selected from the same video. In contrast, the \textbf{inter-video} task uses a pool of 5 clips, each selected from a different video.
% Thus, the \textbf{intra-video} task is more challenging.

\noindent \textbf{Models.}
We use two recent text-audio retrieval models, namely LAION-Clap~\cite{laionclap2023,htsatke2022} and WavCaps~\cite{mei2023wavcaps}. Both models were trained and/or finetuned on the audio-centric AudioCaps~\cite{kim2019audiocaps} and Clotho~\cite{drossos2020clotho} datasets.
We use these models to assess zero-shot egocentric text-audio retrieval capabilities, i.e.\ without any training on egocentric data. \par

% We additionally investigate the impact of the generated audio descriptions on text-video retrieval with audio-visual data. 
% For this, we ensemble the same audio models and a visual model~\cite{lin2022egocentric}. 
% Specifically, for joint evaluation, we generate the required similarity matrices for the audio model and for the video model, and average them before calculating the metrics.\par

\noindent \textbf{Zero-shot egocentric text-audio retrieval.}
% Our primary emphasis is on the EpicMIR dataset, as our methodology for generating audio descriptions aligns with its particular style of verb + noun visual text descriptions. 
% We evaluate the LAION-Clap and WavCaps models in a zero-shot setting for egocentric text-audio retrieval and provide the results on AudioEpicMIR in Tab.~\ref{tab:audio_epicmir_gpt_aug}.
We evaluate the audio-centric pre-trained LAION-Clap and WavCaps models directly on the egocentric text-audio retrieval datasets (zero-shot setting) and provide the results on AudioEpicMIR in Tab.~\ref{tab:audio_epicmir_gpt_aug}.
We observe that the LLM-generated audio descriptions yield a consistent boost.
We hypothesize that this improvement stems both from aligning the style of descriptions more closely with the training distribution for the models and from the inclusion of more audio-centric content.
% The WavCaps model fine-tuned on Clotho (WavCaps-Cl) delivers superior performance to WavCaps-AC and LAION. ChatGPT-generated audio descriptions yield consistently better performance than using the original visual descriptions on AudioEpicMIR. 
We additionally report results for AudioEgoMCQ in Tab.~\ref{tab:audio_egomcq_gpt_aug} \ao{and EpicSoundsRet in Tab.~\ref{tab:audio_epicsounds_gpt_allmodels}}. 
\ao{For AudioEgoMCQ, we observe that for the intra-video task, using LLM descriptions provides consistent improvements over using the original labels. For the inter-video task we observe that the WavCaps model finetuned on Clotho yields a small decrease in performance as compared to using the original labels. We attribute the variation in task performance to the differing text queries and the more visual-centric nature of Clotho descriptions compared to AudioCaps. For EpicSoundsRet, using LLM descriptions gives a significant improvement over using the original audio class labels in most cases.
% This likely explains why the Clotho-finetuned model favors visual labels over audio descriptions in the 'low' subset of AudioEgoMCQ (Tab.~\ref{tab:subset_only_zero_shot_egomcq_aud}). Since this subset constitutes over 50\% of AudioEgoMCQ, it significantly impacts the overall retrieval results shown in 
% Tab.~\ref{tab:audio_egomcq_gpt_aug}
}

\begin{table}
\caption{Zero-shot text-audio retrieval on AudioEpicMIR with LLM-generated audio descriptions compared to using visual descriptions.
WavCaps-AC and WavCaps-Cl were finetuned on AudioCaps and Clotho respectively.}
\vspace{1mm}
\centering
\resizebox{\columnwidth}{!}{%
\begin{tabular}{cccccccc}
 \toprule
 \makecell{Pre-trained \\audio model} & \makecell{LLM-generated\\ audio descriptions} & \multicolumn{3}{c}{mAP(\%)} & \multicolumn{3}{c}{nDCG(\%)} \\
  \cmidrule(lr){3-5} \cmidrule(l){6-8}
 & & A->T & T->A & AVG & A->T & T->A & AVG \\
 \midrule
 \makecell{Random} & & 5.6 & 6.4 & 6.0 & 10.7 & 12.3 & 11.5 \\
  \addlinespace
 LAION-Clap & & 8.9 & 8.4 & 8.6 & 15.3 & 17.0 & 16.2 \\
 LAION-Clap  & \checkmark & \textbf{10.0} & \textbf{9.3} & \textbf{9.6} & \textbf{17.2} & \textbf{18.2} & \textbf{17.7} \\
 \addlinespace
 WavCaps-AC &  & 10.3 & 9.0 & 9.7 & 17.5 & 17.5 & 17.5 \\
%  \hline
   WavCaps-AC & \checkmark & \textbf{11.2} & \textbf{10.4} & \textbf{10.8} & \textbf{18.9} & \textbf{20.0} & \textbf{19.4} \\

 % \hline
 \addlinespace
 WavCaps-Cl  & & 10.9 & 9.5 & 10.2 & 18.3 & 18.2 & 18.2 \\
%   \hline
 WavCaps-Cl  & \checkmark & \textbf{11.5} & \textbf{10.4} & \textbf{10.9} & \textbf{19.2} & \textbf{20.2} & \textbf{19.7} \\

\bottomrule

\end{tabular}%
}
% \vspace*{-6mm}
\label{tab:audio_epicmir_gpt_aug}
\end{table}

\begin{table}[t]
% \vspace*{-5mm}
\caption{Zero-shot text-audio retrieval results on AudioEgoMCQ with LLM-generated audio descriptions compared to using visual descriptions.
\samuel{The data is filtered as per Sec.~\ref{sec:tasks_datasets}.}}
\vspace{1mm}
\centering
\resizebox{\columnwidth}{!}{%
\begin{tabular}{ ccccc }
\toprule
\makecell{Pre-trained \\audio model}& \makecell{LLM-generated\\ audio descriptions} & & Intra-video(\%) & Inter-video(\%) \\
\midrule

Random & & & 20.0 & 20.0 \\
\addlinespace
LAION-Clap & & & 24.2 & 27.5 \\
LAION-Clap & \checkmark & & \textbf{25.2} & \textbf{28.9} \\
\addlinespace
 WavCaps-AC & & & 24.1 & 30.9 \\
 WavCaps-AC & \checkmark & & \textbf{25.1} & \textbf{31.1} \\
 \addlinespace

WavCaps-Cl & & & 24.6 & \textbf{32.1} \\
WavCaps-Cl & \checkmark & & \textbf{25.6} & \underline{31.8} \\
% \addlinespace
%  \hline
% WavCaps-Cl no low & & & 27.9 & 36.2 \\
% WavCaps-Cl no low & \checkmark & & \textbf{29.0} & \textbf{36.9} \\
\bottomrule
\end{tabular}
}
% \vspace*{-2mm}
\label{tab:audio_egomcq_gpt_aug}
\end{table}

\begin{table}
\caption{Zero-shot text-audio retrieval on EpicSoundsRet with LLM-generated audio descriptions compared to using audio class labels.}
\centering
\resizebox{\columnwidth}{!}{%
\begin{tabular}{ccccccccc}
 \toprule
 \makecell{Pre-trained \\audio model} & \makecell{LLM-generated\\ audio descriptions} & \multicolumn{3}{c}{mAP(\%)} & \multicolumn{3}{c}{nDCG(\%)} \\
  \cmidrule(lr){3-5} \cmidrule(l){6-8}
 & & A->T & T->A & AVG & A->T & T->A & AVG \\
 \midrule
   Random   &  & 8.7 & 2.8 & 5.7 & 1.8 & 1.9 & 1.9 \\
\addlinespace
 WavCaps-Cl   &  & 24.3 & \textbf{12.0} & 18.1 & 11.0 & 13.6& 12.3  \\
%   \hline
 WavCaps-Cl  & \checkmark & 30.2 & 11.3 & 20.8 & 14.8 & 13.7 & 14.3  \\

   \addlinespace
 LAION-Clap  &  & 28.5 & 8.2 & 18.3 & 16.1 & 9.5& 12.8  \\
 LAION-Clap   & \checkmark & 29.9 & 11.7 & 20.8 & 16.3 & \textbf{14.3}& 15.3  \\
%   \hline

\addlinespace

 WavCaps-AC  & & 24.3 & 11.9 & 18.2 & 12.4 & 13.8 & 13.1  \\
%   \hline
 WavCaps-AC  & \checkmark & \textbf{31.2} & 11.9 & \textbf{21.5} & \textbf{16.9} & \textbf{14.3} & \textbf{15.6}  \\

\bottomrule

\end{tabular}%
}

\label{tab:audio_epicsounds_gpt_allmodels}
\end{table}

\begin{table}[t]
\caption{Zero-shot text-audio retrieval on different subsets of AudioEpicMIR, split according to the informativeness of audio files as judged by GPT-4. LLM-generated audio descriptions give the best results across all subsets (Aud LLM), compared to using the original visual descriptions (Aud orig).
WavCaps-Cl performs best when the audio files are considered to be highly informative.
}
\centering
\resizebox{\columnwidth}{!}{%
\begin{tabular}{cccccc}
% \begin{tabularx}{\columnwidth}{XXXXXXX}
 \toprule
 \multirow{2}{*}{\makecell{AudioEpicMIR subset}} & \multirow{2}{*}{Descriptions} & \multicolumn{2}{c}{mAP(\%)} & \multicolumn{2}{c}{nDCG(\%)} \\
 \cmidrule(lr){3-4} \cmidrule(l){5-6}
 & & AVG & $\delta$ to rand. perf. & AVG & $\delta$ to rand. perf. \\
 \midrule
 % \makecell{Random\\(from Egovlp)} &  & 5.7 & 5.6 & 5.7 & 10.8 & 10.9 & 10.9 \\
 % \hline

 \multirow{3}{*}{Low} 
 % & \textit{Random perf.}  & 12.9 & - & 24.0 & - \\%\cline{2-6}
  & \textit{Random perf.}  & 12.8 & - & 24.4 & - \\%\cline{2-6}
 
 & Aud orig &  15.5 & 2.7 & 28.2 & 3.8\\%\cline{2-6}

  & Aud LLM &  15.6 & 2.8 & 27.9 & 3.5\\%\cline{2-6}
  
\midrule

 \multirow{3}{*}{Moderate}  & 
 % \textit{Random perf.} & 6.9 & - & 12.1 & - \\%\cline{2-6}
 \textit{Random perf.} & 7.2 & - & 13.0 & - \\%\cline{2-6}

 & Aud orig &  11.5 & 4.3 &  19.1 & 6.1\\%\cline{2-6}

  & Aud LLM & 12.4 & 5.2 & 20.2 & 7.2 \\%\cline{2-6}
  
\midrule

 % \multirow{3}{*}{High}  & \textit{Random perf.} &6.5 & - & 10.3 & - \\%\cline{2-6}
  \multirow{3}{*}{High}  & \textit{Random perf.} &5.7 & - & 9.7 & - \\%\cline{2-6}
 
 & Aud orig & 14.0 & \underline{8.3} & 21.0 & \underline{11.3}\\%\cline{2-6}

  & Aud LLM & 15.2 & \textbf{9.5} & 23.7 & \textbf{14.0}\\%\cline{2-6}
  
\bottomrule
\end{tabular}
% \end{tabularx}
}
% \vspace{-2mm}
\label{tab:subset_only_zero_shot_epicmir}
\end{table}
% On AudioEgoMCQ, the LAION model is better when using ChatGPT-generated audio descriptions compared to the visual descriptions. However, for the WavCaps models, the GPT descriptions do not yield consistent improvements.
%give boosts on the Intra-video task and are a bit worse on the Inter-video task. 
% Nevertheless, the gain when using GPT descriptions on the Intra-video task, which is also the harder task, is higher than the loss on Inter-video.
% We believe that the results would improve with LLM prompts better suited to the style of the EgoMCQ descriptions instead of using a prompt based on EpicMIR. 
%In this particular experiment, the methodology of generating audio descriptions is tailored to very short descriptions since it was developed to make use of the samples that are contained in both Kinetics and AudioCaps.\par

\noindent \textbf{Evaluating text-audio retrieval on subsets with different audio relevancy.}
We employ GPT-4 to split the audio tracks in AudioEpicMIR into three subsets as described in Sec.~\ref{subsec:llmssplits}. The subsets are selected by the LLM based on how difficult it is to identify the audio content solely from the sound.
% These subsets are mostly echilibrated, their sizes for EpicMIR being low: 3544, moderate: 2132, high: 3992. For EgoMCQ the splits are, in the same order, as follows: 13356, 5118 and 6362.
 We compare the performance of different models to a random baseline on these newly created subsets in Tab.~\ref{tab:subset_only_zero_shot_epicmir}.  The random baseline has been obtained by providing randomly sampled text descriptions as the `correct' descriptions for a given audio.
% The random baseline performance was obtained by inputting zeroed audio files and visual descriptions.
\samuel{The audio retrieval model employed is WavCaps-Cl.}
We observe that for the subset deemed to have `low' audio relevance, the random performance more than doubles in comparison to the random evaluation on the full test set.
This is a result of the subset's pool of text descriptions being fairly repetitive.
Hence, to ensure a more equitable comparison of model performance across the subsets, we consider the incremental improvement of each model from its initial random metrics ($\delta$ to rand. perf.).
The accuracy increases as we go from the `low' to the `high' subset. At the same time, the incremental improvement over the random performance is most significant for the 'high' subset.
We perform the same experiment on the AudioEgoMCQ subsets and observe that GPT-4 is indeed capable of selecting videos for which the audio is more likely to have informative content  
% Results are provided in
(Tab.~\ref{tab:subset_only_zero_shot_egomcq_aud}).
% Here we have used the LAION model since it was the one where using ChatGPT generated descriptions improved both metrics of interest.

% \input{tables/egomcq_audio_subsets_only_caption_0sec_laion}
\begin{table}
\caption{
Zero-shot text-audio retrieval on different subsets of AudioEgoMCQ, split according to the informativeness of audio files as judged by GPT-4.
WavCaps-Cl performs best for highly informative audio files. Random values for both tasks are 20.0.}
\vspace{1mm}
\centering
\resizebox{0.85\columnwidth}{!}{%
\begin{tabular}{cccc}
 \toprule
 \makecell{AudioEgoMCQ subset} & Descriptions & Intra-video(\%) & Inter-video(\%) \\
 \midrule
 
 \multirow{2}{*}{Low} 
 
 & Aud orig & 21.7 & 28.2 \\%\cline{2-4}

 & Aud LLM &  22.6 & 26.9 \\%\cline{2-4}
  
\midrule

 \multirow{2}{*}{Moderate} 
 
 & Aud orig & 25.0 & 32.7 \\%\cline{2-4}

 & Aud LLM &  26.8 & 34.2 \\%\cline{2-4}

% \midrule

%  \multirow{2}{*}{Moderate $+$ High} 
 
%  & Aud orig &  27.9 & 36.2 \\

% & Aud LLM &  \textbf{29.1} & \textbf{36.9} \\

\midrule

 \multirow{2}{*}{High} 
 
 & Aud orig &  30.3 & 39.0 \\%\cline{2-4}

& Aud LLM &  \textbf{30.8} & \textbf{39.1} \\%\cline{2-4}

\bottomrule
\end{tabular}%
}

% \vspace{-2mm}
\label{tab:subset_only_zero_shot_egomcq_aud}
\end{table}
% \vspace*{-5pt}

% \noindent \textbf{Multimodal text-video retrieval.}
% Lastly, we show that text-video retrieval benefits from jointly using the audio and visual modalities. We report text-video retrieval results on the EpicMIR task in Tab.~\ref{tab:joint_eval_zero_shot_epicmir}. The joint multimodal evaluation is carried out via late fusion of the outputs of text-audio and text-video retrieval models. 
% Boosts in performance for the multimodal evaluation compared to the unimodal models (i.e.\ audio only and video only) are observed both when using the original descriptions (Joint w.\ Aud orig) and when using the original descriptions for the video model and the LLM-generated ones for the audio model (Joint w.\ Aud LLM).
% \input{tables/joint_eval_zero_shot_epicmir_0sec}
% \vspace*{-30pt}

% \input{tables/epic_audio_subsets_full_caption}
% \input{tables/joint_eval_zeroshot_egomcq}

% \input{tables/joint_epic_audio_subsets_full_caption_0sec}
\vspace{-3mm}
\section{Conclusion}

\label{sec:conclusion}
In this study, we introduced three new benchmarks for egocentric text-audio retrieval. We proposed a methodology of generating audio descriptions using an LLM starting from visual-centric descriptions \ao{and audio class labels}.
% \ao{Furthermore, we show that this methodology out of the box can improve retrieval results when starting from audio-centric labels.}
Lastly, we have shown that we can use guidance from an LLM to filter out noisy audio content extracted from video datasets. We believe these contributions can apply beyond the egocentric setting and hope they will improve text-audio understanding.\par

%\vfill
%\pagebreak

% \section{COPYRIGHT FORMS}
% \label{sec:copyright}

% You must submit your fully completed, signed IEEE electronic copyright release
% form when you submit your paper. We {\bf must} have this form before your paper
% can be published in the proceedings.

\vfill\pagebreak

\bibliographystyle{IEEEbib}
\bibliography{strings,refs}

\begin{thebibliography}{10}

\bibitem{chatgpt}
OpenAI,
\newblock ``Chatgpt,'' \url{https://openai.com/blog/chatgpt},
\newblock Accessed July, August 2023.

\bibitem{2023arXiv230709288T}
H.~{Touvron}, L.~{Martin}, K.~{Stone}, et~al.,
\newblock ``{Llama 2: Open Foundation and Fine-Tuned Chat Models},''
\newblock {\em arXiv:2307.09288}, 2023.

\bibitem{Smaira2020ASN}
L.~Smaira, J.~Carreira, E.~Noland, et~al.,
\newblock ``A short note on the kinetics-700-2020 human action dataset,''
\newblock {\em arXiv:2010.10864}, 2020.

\bibitem{kim2019audiocaps}
C.~D. Kim, B.~Kim, H.~Lee, and G.~Kim,
\newblock ``Audiocaps: Generating captions for audios in the wild,''
\newblock in {\em Proc. NACCL}, 2019.

\bibitem{Damen2018EPICKITCHENS}
D.~Damen, H.~Doughty, G.~M. Farinella, et~al.,
\newblock ``Scaling egocentric vision: The epic-kitchens dataset,''
\newblock in {\em ECCV}, 2018.

\bibitem{ego4d}
K.~Grauman, M.~Wray, A.~Fragomeni, et~al.,
\newblock ``Ego4d: Around the world in 3,000 hours of egocentric video,''
\newblock in {\em CVPR}, 2022.

\bibitem{Oncescu21a}
A.-M. Oncescu, {\relax A. S}.~Koepke, J.~Henriques, et~al.,
\newblock ``Audio retrieval with natural language queries,''
\newblock in {\em INTERSPEECH}, 2021.

\bibitem{koepke2022audio}
A.~S. Koepke, A.-M. Oncescu, J.~Henriques, et~al.,
\newblock ``Audio retrieval with natural language queries: A benchmark study,''
\newblock {\em IEEE Transactions on Multimedia}, 2022.

\bibitem{slaney2002semantic}
M.~Slaney,
\newblock ``Semantic-audio retrieval,''
\newblock in {\em ICASSP}, 2002.

\bibitem{laionclap2023}
Y.~Wu, K.~Chen, T.~Zhang, et~al.,
\newblock ``Large-scale contrastive language-audio pretraining with feature fusion and keyword-to-caption augmentation,''
\newblock in {\em ICASSP}, 2023.

\bibitem{OpenAI2023GPT4TR}
OpenAI,
\newblock ``Gpt-4 technical report,''
\newblock {\em arXiv:2303.08774}, 2023.

\bibitem{vicuna2023}
W.-L. Chiang, Z.~Li, Z.~Lin, et~al.,
\newblock ``Vicuna: An open-source chatbot impressing gpt-4 with 90\%* chatgpt quality,'' 2023.

\bibitem{touvron2023llama}
H.~Touvron, T.~Lavril, G.~Izacard, et~al.,
\newblock ``Llama: Open and efficient foundation language models,''
\newblock {\em arXiv:2302.13971}, 2023.

\bibitem{zhu2023minigpt}
D.~Zhu, J.~Chen, X.~Shen, et~al.,
\newblock ``Minigpt-4: Enhancing vision-language understanding with advanced large language models,''
\newblock {\em arXiv:2304.10592}, 2023.

\bibitem{shen2023hugginggpt}
Y.~Shen, K.~Song, X.~Tan, et~al.,
\newblock ``Hugginggpt: Solving ai tasks with chatgpt and its friends in huggingface,''
\newblock {\em arXiv:2303.17580}, 2023.

\bibitem{Kaul23}
P.~Kaul, W.~Xie, and A.~Zisserman,
\newblock ``Multi-modal classifiers for open-vocabulary object detection,''
\newblock in {\em ICML}, 2023.

\bibitem{menon2022visual}
S.~Menon and C.~Vondrick,
\newblock ``Visual classification via description from large language models,''
\newblock in {\em ICLR}, 2023.

\bibitem{mei2023wavcaps}
X.~Mei, C.~Meng, H.~Liu, et~al.,
\newblock ``Wavcaps: A chatgpt-assisted weakly-labelled audio captioning dataset for audio-language multimodal research,''
\newblock {\em arXiv:2303.17395}, 2023.

\bibitem{NEURIPS2022_96471570}
H.~Mittal, P.~Morgado, U.~Jain, and A.~Gupta,
\newblock ``Learning state-aware visual representations from audible interactions,''
\newblock in {\em NeurIPS}, 2022.

\bibitem{Planamente_2022_WACV}
M.~Planamente, C.~Plizzari, E.~Alberti, and B.~Caputo,
\newblock ``Domain generalization through audio-visual relative norm alignment in first person action recognition,''
\newblock in {\em WACV}, 2022.

\bibitem{lin2022egocentric}
K.~Q. Lin, J.~Wang, M.~Soldan, et~al.,
\newblock ``Egocentric video-language pretraining,''
\newblock in {\em NeurIPS}, 2022.

\bibitem{zhao2023lavila}
Y.~Zhao, I.~Misra, P.~Kr{\"a}henb{\"u}hl, and R.~Girdhar,
\newblock ``Learning video representations from large language models,''
\newblock in {\em CVPR}, 2023.

\bibitem{Ashutosh_2023_CVPR}
K.~Ashutosh, R.~Girdhar, L.~Torresani, and K.~Grauman,
\newblock ``Hiervl: Learning hierarchical video-language embeddings,''
\newblock in {\em CVPR}, 2023.

\bibitem{pramanick2023egovlpv2}
S.~Pramanick, Y.~Song, S.~Nag, et~al.,
\newblock ``Egovlpv2: Egocentric video-language pre-training with fusion in the backbone,''
\newblock in {\em ICCV}, 2023.

\bibitem{Kazakos2019EPICFusionAT}
E.~Kazakos, A.~Nagrani, A.~Zisserman, and D.~Damen,
\newblock ``Epic-fusion: Audio-visual temporal binding for egocentric action recognition,''
\newblock in {\em ICCV}, 2019.

\bibitem{EPICSOUNDS2023}
J.~Huh, J.~Chalk, E.~Kazakos, et~al.,
\newblock ``{EPIC-SOUNDS}: {A} {L}arge-{S}cale {D}ataset of {A}ctions that {S}ound,''
\newblock in {\em ICASSP}, 2023.

\bibitem{damen2022rescaling}
D.~Damen, H.~Doughty, G.~M. Farinella, et~al.,
\newblock ``Rescaling egocentric vision,''
\newblock {\em IJCV}, 2022.

\bibitem{10.1145/582415.582418}
K.~J\"{a}rvelin and J.~Kek\"{a}l\"{a}inen,
\newblock ``Cumulated gain-based evaluation of ir techniques,''
\newblock {\em ACM Trans. Inf. Syst.}, 2002.

\bibitem{htsatke2022}
K.~Chen, X.~Du, B.~Zhu, et~al.,
\newblock ``Hts-at: A hierarchical token-semantic audio transformer for sound classification and detection,''
\newblock in {\em ICASSP}, 2022.

\bibitem{drossos2020clotho}
K.~Drossos, S.~Lipping, and T.~Virtanen,
\newblock ``Clotho: An audio captioning dataset,''
\newblock in {\em ICASSP}, 2020.

\bibitem{wray2019fine}
M.~Wray, D.~Larlus, G.~Csurka, and D.~Damen,
\newblock ``Fine-grained action retrieval through multiple parts-of-speech embeddings,''
\newblock in {\em ICCV}, 2019.

\bibitem{EgoVLP2023code}
{Show Lab},
\newblock ``Egovlp: A repository for ego-centric vision language pre-training,'' \url{https://github.com/showlab/EgoVLP}, 2023,
\newblock Accessed: 2023-01-10.

\end{thebibliography}
\clearpage
\appendix
\section{Supplementary Material:\\ A SOUND APPROACH: USING LARGE LANGUAGE MODELS TO GENERATE AUDIO
DESCRIPTIONS FOR EGOCENTRIC TEXT-AUDIO RETRIEVAL}
In this supplementary material, we will provide further experimental results and details regarding the pre-trained models that we used. In Sec.~\ref{sec:epicsounds}, we present additional information and analysis of the EpicSoundsRet benchmark based on the EpicSounds~\cite{EPICSOUNDS2023} dataset. For completeness, we include the prompts for generating audio descriptions in Sec.~\ref{sec:prompts}. In Sec.~\ref{sec:video_model}, we evaluate the impact of additionally using the audio modality for the text-video retrieval task. We provide more insight into the differences between the Clotho and AudioCaps finetuned checkpoints used in our work in Sec.~\ref{sec:clotho_vs_audiocaps}. In Sec.~\ref{sec:intra_vs_inter} we briefly expand on the results obtained on the AudioEgoMCQ task whilst in Sec.~\ref{sec:audioegomcq_audiocaps_extra} we provide more experiments on the AudioEgoMCQ benchmark. Lastly, we compare the relative performance differences of the models used for the experiments in the paper on multiple benchmarks in Sec.~\ref{sec:cl_vs_ac_on_vis_vs_aud}.

\subsection{EpicSounds}\label{sec:epicsounds}
In our experiments with the AudioEpicMIR and AudioEgoMCQ benchmarks, we generated \textit{audio descriptions} using LLMs starting from \textit{visual} class labels. 
For EpicSounds, we have access to \textit{audio} class labels. As a result, our prompts are \textit{not} necessarily optimal when using \textit{audio} class labels as inputs. Nevertheless, despite being created from a set of prompts that might not be ideal, the \textit{audio descriptions} still perform better in the retrieval task compared to the original audio labels as can be seen in Tab.~\ref{tab:audio_epicsounds_gpt_allmodels} in the main paper. 

The WavCaps model finetuned on Clotho performs slightly worse on the text-audio task when using LLM-generated descriptions on EpicSounds than when using the original audio labels. However, we do not see this behaviour when using models finetuned on AudioCaps. This discrepancy likely arises because Clotho descriptions often include substantial visual details in addition to audio content, leading to a significant style difference in our audio-focused descriptions. However, regardless of the model used, LLM-generated descriptions overall yield better results than using class labels, particularly in the audio-to-text retrieval direction.\par

We employed the same retrieval metrics for AudioEpicMIR and EpicSoundsRet. This required constructing a relevancy matrix (more information on this is available here~\cite{wray2019fine, damen2022rescaling}). This relevancy matrix takes into account that for one given description, multiple audio files can be equally relevant. On AudioEpicMIR, the relevancy matrix would look at all unique text descriptions, and depending on how many nouns and verbs the other descriptions would have in common, it would assign values between 0 and 1 to these similar descriptions. For EpicSoundsRet we start with audio class labels as `descriptions' and assign a score of 1 to all audios and text descriptions where the text description belongs to the same class. We use the same relevancy matrix to calculate the retrieval scores for the \textit{Aud orig} and \textit{Aud LLM} descriptions. This also applies to AudioEpicMIR.\par
% On all the other tasks and models, using LLM generated audio descriptions outperforms using audio labels for retrieval.
% Lastly, these results are obtained by using prompts that were not tailored for taking in audio labels. We hypothesise there is room for improvement in terms of prompt engineering that can improve the numbers further.
% \input{tables/audio_eval_epicsounds_zeroshot_gpt_0sec_allmodels}

\subsection{Prompts}\label{sec:prompts}
We use LLMs, specifically GPT3.5, to generate audio descriptions starting from visual class labels. Tab.~\ref{tab:prompts_table} shows the prompts we used.\par
\begin{table*}
\caption{Overview of the methodology for bridging the gap between visual classes/descriptions and video soundtrack descriptions using LLMs. The process includes setting the scene, few-shot prompting with examples, and generating audio descriptions from video descriptions.}
\centering
% \resizebox{\columnwidth}{!}{%
% \begin{tabular}{cccccc}
\begin{tabular}{>{\centering\arraybackslash}p{0.5in} p{6in}}
 \toprule
 \multicolumn{1}{c}{Index} & \multicolumn{1}{c}{Prompts} \\
\midrule
  1. &You are an expert in audio and visual description of videos. You have seen the AudioCaps and Clotho datasets and know how to generate relevant audio descriptions using simple terms. You will help me generate audio descriptions that can match the audio content of a video for which the video description is provided. Avoid the use of object or actions that cannot be inferred from the audio signal alone. Try to create proper short sentences when  generating the audio descriptions. When multiple audio descriptions can be possible, provide one that best generalises the sounds.Try to avoid generating audio descriptions in the form of 'sound of [visual object/action]' and instead use the actual noise or sound that object/action can make. If unsure, you can provide multiple possible sounds. To help you understand the sort of descriptions I am looking for, in the next prompt I will provide you a few pairs of video descriptions and the corresponding audio descriptions as an example. Please remember this prompt and the examples whenever generating new audio descriptions. Then I will ask you to generate audio descriptions given new video descriptions. \\
  \addlinespace
  2. &Here are some examples in the form of (video description: audio description) and examples are separated by semicolon.
('burping', 'A man giving out a loud burp');('sneezing', 'Someone sneezes');('washing dishes', 'Metal clinking and clanging occur');('washing dishes', 'Water splashing and glasses clanging together then more clanging ending with glass squeaking sound');('opening door', 'Door handle continuously clicking then being pushed open');('spray painting', 'Powerful bursts of spraying');('crying', 'A baby cries and screams');('opening door', 'Hissing then creaking as a door is opened');('applauding', 'A large number of people clap, cheer, and shout');('closing door', 'Metal clings followed by rumbling and metal sliding');('whistling', 'A person is whistling a tune');('shearing sheep', 'Sheep bleat quietly');('washing hair', 'Water running while the stream is interrupted at times');('sawing wood', 'Rubbing and sawing of wood'). I want you to learn from them and not generate descriptions for this prompt.\\
\addlinespace
3. &Generate an audio description for each of the following enumerated video descriptions separated by semicolon. Try to avoid generating audio descriptions in the form of 'sound of [visual object/action]' and instead use the actual noise or sound that object/action can make. If unsure, you can provide multiple possible sounds. Remember the original prompt. Provide your answers in the form [description index. video description: generated audio description].\newline
<Add the list of visual labels separated by ; here>\\

\bottomrule

\end{tabular}%
% }
\label{tab:prompts_table}
\end{table*}

Furthermore, we used GPT4 to decide the difficulty of identifying the action associated with a sound. The prompt we used for that is provided in Tab.~\ref{tab:audio_relevance}.

\begin{table*}
\caption{Prompts used for employing LLMs in assessing the likelihood of identifying video actions solely through audio tracks.}
\centering
% \resizebox{\columnwidth}{!}{%
% \begin{tabular}{cccccc}
\begin{tabular}{>{\centering\arraybackslash}p{0.5in} p{6in}}
 \toprule
 \multicolumn{1}{c}{Index} & \multicolumn{1}{c}{Prompts} \\
\midrule
  1. &You have a lot of experience with video and audio descriptions. You are working with videos that have an associate audio file. You only have descriptions of the visual content. Some videos are highly correlated with the audio content, such as those depicting someone cutting vegetables. Other videos have very generic audios and if only given the audio content, the video content might not be easy to figure out. I want you to tell me if a video description is relevant for the possible associated audio content. I want you to provide your answers in the form of a dictionary where the key is the description I am giving you and the value is the relevance. Relevance should be high if the sound associated is easy to assume. Relevance should be moderate if the associated sounds can be more than one. Relevance should be low if it's unlikely to hear any specific sounds. Process each entry individually. Input descriptions will be given in the form of a list. Do not provide additional comments, just the relevance. \\

\bottomrule

\end{tabular}%
% }

\label{tab:audio_relevance}
\end{table*}

\subsection{Multimodal text-video retrieval}\label{sec:video_model}

In this section, we show that text-video retrieval benefits from jointly using the audio and visual modalities. We report text-video retrieval results on the EpicMIR task in Tab.~\ref{tab:joint_eval_zero_shot_epicmir}. The joint multimodal evaluation is carried out via late fusion of the outputs of the text-audio and text-video retrieval models. 
Boosts in performance for the multimodal evaluation compared to the unimodal models (i.e.\ audio only and visual only) are observed both when using the original descriptions (Joint w.\ Aud orig) and when using the original descriptions for the visual model together with the LLM-generated ones for the audio model (Joint w.\ Aud LLM). \par
\begin{table}[h!]
\caption{Zero-shot capabilities of text-audio (audio only) and text-video (video only) retrieval models compared to the joint evaluation with late fusion on EpicMIR. $^\ast$ Numbers are obtained using the 
text-video retrieval model from \cite{lin2022egocentric}.
}
\vspace{1mm}
\centering
\resizebox{\columnwidth}{!}{%
\begin{tabular}{ccccccc}
\toprule
\multirow{2}{*}{EpicMIR} & \multicolumn{3}{c}{mAP(\%)} & \multicolumn{3}{c}{nDCG(\%)} \\
\cmidrule(lr){2-4} \cmidrule(l){5-7}
 & A->T & T->A & AVG & A->T & T->A & AVG \\
\midrule
Random~\cite{lin2022egocentric}  & 5.7 & 5.6 & 5.7 & 10.8 & 10.9 & 10.9 \\
\addlinespace
 Audio only (WavCaps-Cl w.\ Aud LLM) & 11.5 & 10.5 & 11.0 & 19.2 & 20.3 & 19.7 \\
\addlinespace
Video only (EgoVLP$^\ast$) & 24.7 & 18.4 & 21.6 & 27.4 & 24.6 & 26.0 \\
\addlinespace
 Joint w.\ Aud orig & \underline{25.4} & \underline{19.3} & \underline{22.4} & \underline{28.7} & \underline{26.0} & \underline{27.3} \\
\addlinespace
Joint w.\ Aud LLM & \textbf{25.8} & \textbf{19.9} & \textbf{22.8} & \textbf{29.1} & \textbf{26.9} & \textbf{28.0} \\

\bottomrule
\end{tabular}
}

% \vspace*{-10pt}
\label{tab:joint_eval_zero_shot_epicmir}
\end{table}
Additionally, we investigate how text-video retrieval performs on the same audio relevancy subsets investigated in Tab.~\ref{tab:subset_only_zero_shot_epicmir}. Results are provided in Tab.~\ref{tab:subset_only_zero_shot_epicmir_audvid} and follow a similar trend as for the audio only experiments. More specifically, the multimodal text-video retrieval performs best on the `high' subset. We also notice that text-video retrieval when only using the visual modality also exhibits stronger performance on subsets with higher audio relevancy. We hypothesise that this is because the more audio significant actions tend to also be more common, or easier to identify for visual models.
\begin{table}[t]
\caption{Zero-shot multimodal text-video retrieval on different subsets of AudioEpicMIR, split according to the informativeness of audio files as judged by GPT-4. LLM-generated audio descriptions give the best results across all subsets (Aud LLM), compared to using the original visual descriptions (Aud orig).
The joint model performs best when the audio files are considered to be highly informative.
}
\centering
\resizebox{\columnwidth}{!}{%
\begin{tabular}{cccccc}
% \begin{tabularx}{\columnwidth}{XXXXXXX}
 \toprule
 \multirow{2}{*}{\makecell{AudioEpicMIR subset}} & \multirow{2}{*}{Descriptions} & \multicolumn{2}{c}{mAP(\%)} & \multicolumn{2}{c}{nDCG(\%)} \\
 \cmidrule(lr){3-4} \cmidrule(l){5-6}
 & & AVG & $\delta$ to rand. perf. & AVG & $\delta$ to rand. perf. \\
 \midrule
 % \makecell{Random\\(from Egovlp)} &  & 5.7 & 5.6 & 5.7 & 10.8 & 10.9 & 10.9 \\
 % \hline

 \multirow{4}{*}{Low} 
 % & \textit{Random perf.}  & 12.9 & - & 24.0 & - \\%\cline{2-6}
  & \textit{Random perf. vid.}  & 12.5 & - & 23.4 & - \\%\cline{2-6}
  & Vid orig & 23.3 & 10.8 & 30.8 & 7.4\\%\cline{2-6}
 & Aud orig &  23.7 & 11.2 & 32.1 & 8.7\\%\cline{2-6}

  & Aud LLM & 23.4  & 10.9 & 32.1 & 8.7\\%\cline{2-6}
  
\midrule

 \multirow{4}{*}{Moderate}  & 
 % \textit{Random perf.} & 6.9 & - & 12.1 & - \\%\cline{2-6}
 \textit{Random perf. vid.} & 7.2 & - & 12.2 & - \\%\cline{2-6}
 
  & Vid orig & 27.3 & 20.1 & 28.0 & 15.8\\%\cline{2-6}
 & Aud orig & 27.7 & 20.5 & 29.1 & 16.9\\%\cline{2-6}

  & Aud LLM & 28.5 & 21.3 & 29.7 & 17.5 \\%\cline{2-6}
  
\midrule

 % \multirow{3}{*}{High}  & \textit{Random perf.} &6.5 & - & 10.3 & - \\%\cline{2-6}
  \multirow{4}{*}{High}  & \textit{Random perf. vid.} &5.8& - & 9.1 & - \\%\cline{2-6}
  & Vid orig & 32.9 & 27.1 & 36.0 & 26.9\\%\cline{2-6}
 & Aud orig & 33.6 &27.8& 37.1 & 28.0\\%\cline{2-6}

  & Aud LLM & 34.6 &\textbf{28.8}& 38.3 & \textbf{29.2}\\%\cline{2-6}
  
\bottomrule
\end{tabular}
% \end{tabularx}
}
% \vspace{-2mm}
\label{tab:subset_only_zero_shot_epicmir_audvid}
\end{table}

For the text-video retrieval experiments, we used the EgoVLP~\cite{lin2022egocentric} codebase. More precisely, we take the provided pre-trained EgoVLP~\cite{EgoVLP2023code} model and evaluate it in a zero-shot fashion. This model processes the clips by randomly selecting a number of frames. We set the number of frames to 16 in our experiments.
% Additionally, we observe that depending on how the opencv~\cite{itseez2014theopencv,itseez2015opencv} library is called when reading the frames in this codebase, the retrieval results vary by 3-4\%. In this work, we have opted for the top (green) image processing approach that gives the best performance. We believe that the bottom approach (red) uses a wrong flag which was used in the original experiments of the EgoVLP paper. 
Additionally, we do not use the dual\_softmax evaluation approach used in the codebase~\cite{EgoVLP2023code} but instead leverage the standard similarity matrix at evaluation time. This is consistent with how the underlying EgoVLP model was trained~\cite{EgoVLP2023code}.

% \begin{lstlisting}[language=Python,breaklines=true,identifierstyle=\color{green!40!black}]
% frame = cv2.imread(os.path.join(video_path, img_name))
% frame = cv2.cvtColor(frame, cv2.COLOR_BGR2RGB)

% \end{lstlisting}

% \begin{lstlisting}[language=Python,frame=single,breaklines=true,identifierstyle=\color{red}]
% frame = cv2.imread(os.path.join(video_path, img_name), cv2.COLOR_BGR2RGB)

% \end{lstlisting}

\subsection{Leveraging models finetuned on Clotho for tasks with original visual descriptions}\label{sec:clotho_vs_audiocaps}
The Clotho~\cite{drossos2020clotho} dataset was collected by providing annotators with audio files and asking them to generate descriptions of the audio sound. However, by looking at descriptions, we observe that the annotators tended to provide plausible visual descriptions of the sounds rather than describe the actual sound. Some examples are ``A group is in a carriage that is being drawn by a horse on a paved road." or ``A man moves from the basement to upstairs, moving a heavy metal object with him.". Such visual details cannot be inferred just by listening to an audio. Therefore, this dataset matches better our setting of generating audio descriptions starting from visual descriptions since, when no obvious audio description can be generated, the output description often contains a reworded version of the provided \textit{visual} input. In contrast, AudioCaps descriptions are shorter and more audio focused e.g. ``Speech in the distance with a bleating sheep nearby." or ``Food and oil sizzling followed by a woman speaking.". 
Furthermore, the WavCaps model has been finetuned on the Clotho data, which contains 25 times fewer training examples than AudioCaps. This could mean that the model finetuned on Clotho retains more  generality which can be useful in settings with new \textit{audio} content that the LLM might generate.%\end{itemize}
\par

\subsection{Intra-video vs inter-video results}\label{sec:intra_vs_inter}

When using the WavCaps-Cl model, we notice in Tab.~\ref{tab:audio_egomcq_gpt_aug} that the inter-video task performs slightly better when using original descriptions compared to using the LLM-generated ones. Based on our analysis, we believe that this is due to the different distribution of text queries between the two tasks (i.e.\ intra-video and inter-video). 
\subsection{Evaluation on different subsets of AudioEgoMCQ according to informativeness}\label{sec:audioegomcq_audiocaps_extra}
We additionally evaluate the WavCaps model finetuned on AudioCaps on the \textit{low}, \textit{moderate}, and \textit{high} subsets of AudioEgoMCQ in Tab.~\ref{tab:subset_only_zero_shot_egomcq_aud_wavcapsac}. We notice that when using this checkpoint, the inter-video LLM performance is higher on the \textit{moderate} and \textit{high} and just a bit lower on the \textit{low} subset. This is in accordance with results in Tab.~\ref{tab:subset_only_zero_shot_egomcq_aud} where the checkpoint used was finetuned on Clotho. However, when using the AudioCaps finetuned model with LLM descriptions, the decrease on the \textit{low} subset is much lower than when using the Clotho finetuned model. We believe that this is related to Clotho being more visual based as described  in~\ref{sec:clotho_vs_audiocaps}. As a result, when using AudioCaps finetuned models, the overall retrieval performance is better for the LLM descriptions than when using Clotho based models.

\begin{table}
\caption{
Zero-shot text-audio retrieval on different subsets of AudioEgoMCQ, split according to the informativeness of audio files as judged by GPT-4.
WavCaps-AC performs best for highly informative audio files. Random values for both tasks are $20.0$.}
\vspace{1mm}
\centering
\resizebox{0.85\columnwidth}{!}{%
\begin{tabular}{cccc}
 \toprule
 \makecell{AudioEgoMCQ subset} & Descriptions & Intra-video(\%) & Inter-video(\%) \\
 \midrule
 
 \multirow{2}{*}{Low} 
 
 & Aud orig & 21.0 & 27.0 \\%\cline{2-4}

 &Aud LLM &  22.4 & 26.9 \\%\cline{2-4}
  
\midrule

 \multirow{2}{*}{Moderate} 
 
 & Aud orig & 26.2 & 32.7 \\%\cline{2-4}

 & Aud LLM &  26.6 & 33.2 \\%\cline{2-4}

% \midrule

%  \multirow{2}{*}{Moderate $+$ High} 
 
%  & Aud orig &  27.9 & 36.2 \\

% & Aud LLM &  \textbf{29.1} & \textbf{36.9} \\

\midrule

 \multirow{2}{*}{High} 
 
 & Aud orig &  28.7 & 37.0 \\%\cline{2-4}

& Aud LLM &  \textbf{29.1} & \textbf{37.4} \\%\cline{2-4}

\bottomrule
\end{tabular}%
}

% \vspace{-2mm}
\label{tab:subset_only_zero_shot_egomcq_aud_wavcapsac}
\end{table}

\subsection{Why does WavCaps-Cl give better results on AudioEpicMIR and AudioEgoMCQ whilst the WavCaps-AC model is better on EpicSoundsRet?}\label{sec:cl_vs_ac_on_vis_vs_aud}
We found that the WavCaps model, when finetuned with AudioCaps, performs best for EpicSounds. In contrast, for the other two datasets, the top-performing model is WavCaps finetuned on Clotho. This variation in performance is linked to the nature of the input labels provided to the LLM. Specifically, EpicSounds uses only audio inputs, while AudioEpicMIR and AudioEgoMCQ rely solely on visual inputs. LLMs tend to merge the visual and audio information in the same sentence, especially when the input leans more towards visual content. As AudioCaps descriptions primarily focus on audio, whereas Clotho provides a balanced mix of both audio and visual details. Tasks that are more audio-centric benefit significantly from models finetuned on AudioCaps. Conversely, tasks with a visual emphasis favour models finetuned on Clotho, due to their original labels being more visual-centric.\par

\end{document}